\documentclass[twocolumn, astrosymb, tighten]{aastex631}
\usepackage{derivative}
\usepackage{xcolor}
\usepackage{soul}
\usepackage{mathtools}  
\usepackage{amsmath}
\usepackage{amssymb}
\usepackage{xfrac}
\usepackage{float}
\usepackage{graphicx}
\usepackage{placeins}
\usepackage{txfonts}
\begin{document}
\title{DYNAMICAL EVOLUTION OF SECOND-GENERATION CIRCUMSTELLAR/PROTOPLANETARY DISKS \\
IN EVOLVED WIDE BINARY SYSTEMS}
\author{Raffaele S. Cattolico}
\affiliation{Technion-Israel Institute of Technology, Haifa, Israel}
\author{Hagai B. Perets}
\affiliation{Technion-Israel Institute of Technology, Haifa, Israel}
\begin{abstract}
In mass-transferring wide binary stellar systems, the companion star can capture some of the mass released in wind by the primary evolved star, and form an accretion disk.
Such accretion disks could evolve to form disks of comparable properties to protoplanetary disks and may enable the formation of new planets and/or the interactions, re-growth, and re-migration of pre-existing planets in the newly formed disks. 
We study the formation and the dynamical evolution of such "second generation" (SG) protoplanetary disks in  evolved wide binary systems, with primaries in the mass range of 1-8 Solar mass. We follow their evolution from the asymptotic giant branch (AGB) phase of the red giant stellar donor until its evolution to become a white dwarf. 
We perform 1D semi-analytical numerical simulations for several binary systems varying the mass of the evolved stellar donor and the initial orbital distance, taking into account the changing mass-loss rates and the binary orbital expansion due to mass-loss.
We calculate the radial density $\Sigma(t,r)$ profile of the formed SG accretion disk and its temperature profile $T(t,r)$, considering  a non-stationary viscosity profile $\nu(t,r)$ which depends on the radial temperature profile.
We find that SG circumstellar disks evolve to form a long-lived stable structure over the lifetime of the donor star, and we show that we can consistently produce the observed conditions and accretion rate inferred for the Mira evolved wide-binary system. More generally, we find that in most cases, the final total masses of SG disks exceed the Minimum Mass Solar Nebula. The quasi-steady state radial surface density profiles are comparable with the typical range of masses and densities of observed (regular "first-generation") protoplanetary disks. This suggests that realistic SG disks can give rise to a second phase of planet formation and dynamics in old wide binary systems. We provide a brief discussion of the observational consequences.
\end{abstract}

\keywords{accretion, accretion disks – binaries: symbiotic – protoplanetary disks – stars: AGB and post-AGB – stars: winds}

\footnote{Corresponding author: raffaelec@campus.technion.ac.i} 

\section{Introduction} \label{sec:intro}

The mass transfer from an evolved stellar donor to its stellar companion represents a typical result of stellar binary evolution. 
In this study, we investigate wide binary systems with orbital separations larger than $10$ AU, where the primary star is an AGB donor star, which experiences mass loss through a stellar wind. Such wide systems allow for the formation of a circumstellar accretion disk around the accreting companion.

In these systems, a fraction of the stellar wind is captured by the companion (where in our current study we mostly consider main sequence solar-type stellar companions).

The disks formed through such mass-transfer evolution are essentially viscous accretion disks, whose lifetime is mostly determined by  the timescale of the asymptotic giant branch (AGB) phase of the stellar donor (\cite{2013ApJ...764..169P}).
The lifetime of such disks can therefore be long, but they become relatively massive only during the increased mass-loss phases in the AGB phase, which is our main focus here.

\cite{per10a} and \cite{2013ApJ...764..169P} suggested that such disks could have properties similar to protoplanetary disks and may allow for a renewed epoch of planet  formation and/or the interaction, growth, and migration of pre-existing planets inside the newly formed "second generation" (SG) disks. If the secondary later evolves too, to donate mass to the, now, white dwarf (WD) original primary, a SG protoplanetary disk may form around a WD.  Here we explore the properties of SG disks in realistic evolved wide binary systems, mostly considering main-sequence secondaries, but also exploring the case of the Mira system, where the secondary accretor is already a WD. 

Previous studies of low-mass protoplanetary circumstellar gaseous disks have used several disk models at different levels of complexity (e.g. \cite{2004ApJ...604..388I}; \cite{McNeil_2005}; \cite{2009A&A...501.1161M}; \cite{Chambers_2016}). 
The simplest case considered is a pure gaseous disk with fixed surface density and temperature profiles, parameterized as power laws.

More realistic models considered surface density profiles that decrease over time, including self-similar solutions for viscously evolving disks developed by \cite{lyn+74}.

More complex models calculate the temperature and surface density evolution self-consistently taking into account heating by viscous accretion and irradiation by the central star (\cite{1997ApJ...486..372B}; \cite{2013A&A...549A.124B}). 
SG circumstellar disks in binaries have been studied using several different approaches, both analytical and numerical (\cite{mas+98}; \cite{dev+09}).
\cite{2013ApJ...764..169P} (Hereafter PK13) studied the evolution of non-stationary disks and have shown that disks in binaries, with a constant separation less than 100 AU and with a constant mass of the donor and the accretor, reach a steady state in $ \sim 1$ Myr. 
In this work, we extend the study by PK13, but considering more realistic systems, as well as a wider range of the donor masses and initial orbital separations $a$. In particular, we consider realistic mass-loss histories arising from detailed stellar evolution models (from MESA: \cite{2011ApJS..192....3P}), and account for the dynamical variation of the semi-major axis, $a$, of the wide binary, due to the decreasing of the mass of the stellar donor (which, in turn, affects the mass capture rate by the companion).
As we discuss below, our main findings are the reproduction of the observed Mira system inferred accretion rates and conditions and showing that realistic evolved wide-binary systems produce SG disks with properties very similar to those of regular ("first generation") protoplanetary disks in young environments, and hence provide a fertile ground for a second epoch of planet formation, migration, and growth even in very old systems.

This paper is organized as follows. 
In Section 2, we explain the formation processes of second-generation circumstellar disks. In Section 3, we describe our model and the assumptions involved. 
We then present and discuss our findings in Section 4, and summarize our study in Section 5.
\section{Second generation circumstellar disks}
During the thermally pulsing asymptotic giant branch phase, low and intermediate-mass stars develop strong mass loss and eject their envelope to eventually become white dwarfs.
Stellar winds serve as key ingredients that contribute to the chemical evolution of the Galaxy, releasing the products of stellar nucleosynthesis into the interstellar medium.
In binary systems, the interaction of the wind of the evolved star with its companion can lead to the formation of circumstellar structures, potentially having similar properties to regular protoplanetary disks.
In addition, such mass transfer processes can also affect the composition of the companion star as it  accretes processed wind material lost by the evolved primary. 

In order to model these processes we begin by following \cite{sok+00} to study the conditions for the formation of circumstellar disks from wind accreted material.
We define $j_a$ as the specific angular momentum of the accreted material and $j_2$ as the specific angular momentum of a particle in a Keplerian orbit at the equator of an accreting star.
The formation of a circumstellar structure, by means of an exogenous source of mass, occurs whenever the following condition is satisfied:
\begin{equation} \label{Soker}
j_a > j_2
\end{equation}
In the case of accretion from a stellar wind, the specific angular momentum of the material entering in the Bondi-Hoyle accretion radius (\cite{wan81}) is:
\begin{equation}
    j_a = 
    \eta \
    0.5
    \left(\frac{2 \pi}{P}\right)
    {R_{a}}^2
\end{equation}
with $P$ the orbital period of the material inside the Bondi-Hoyle accretion radius and $\eta$ the Mach number for an isothermal flow ($\eta \sim 0.1$) or an adiabatic flow ($\eta \sim 0.3$).
The specific angular momentum of a particle in a Keplerian orbit is:
\begin{equation}
    j_2 = \sqrt{G M_2 R_2}
\end{equation}
with $M_2$ the mass and $R_2$ the radius of the accreting star.
\newline
We can arrange the condition as follow:
\begin{equation}
      \begin{aligned}
        1 < \frac{j_a}{j_2}
        \simeq
        1.2
        \left(\frac{\eta}{0.2}\right)
        \left(\frac{M_1 + M_2}{2.5 M_{\odot}}\right)
        \left(\frac{M_2}{M_{\odot}}\right)^{\frac{3}{2}} 
        \qquad \qquad
        \\
        \times \left(\frac{R_2}{R_{\odot}}\right)^{-\frac{1}{2}} 
        \left(\frac{a}{100 AU}\right)^{-\frac{3}{2}}
        \left(\frac{v_{r}}{10 km s^{-1}}\right)^{-4}
  \end{aligned}
\end{equation}
Where $M_1$ is the donor stellar mass, $M_2$ the accretor stellar mass, $a$ the semi-major axis of the binary system, and $v_{r}$ is the relative velocity between the wind and the stellar accretor defined as:
\begin{equation}\label{eq:v_r}
    v_{r}=\sqrt{{v_{w}}^{2} + {v_{s}}^{2}, }
\end{equation}
where $v_{w}$ is the velocity of the wind at the location of the secondary star and $v_{s}$ the orbital velocity of the secondary star:
\begin{equation}\label{eq:v_s}
v_{s} = \sqrt{\frac{G {M_{1}}^{2}}{(M_{1} + M_{2}) \ a(t)}}
\end{equation}
Typical values of wind velocity ($v_w$)  are in the range  $5 - 30$ km s$^{-1}$ (\cite{refId0}, \cite{refId01}, \cite{2018A&ARv..26....1H}, \cite{2022IAUS..366..165H}).
In this work, we set a constant value $v_{w} = 10$ km s$^{-1}$. 

\subsection{The mass-loss and accretion rate}
The mass-loss rates $\dot{M}_{w}$ from the stellar donor, in our models, are calculated based on simplified stellar evolution prescriptions  of the single-star evolution \textit{SSE} model (\cite{2000MNRAS.315..543H}).
We assume an isotropic stellar wind from the donor star. Therefore, the companion can capture only a fraction of the stellar wind and form an accretion disk. Thus, the accretion rate onto the disc is:
\begin{equation} \label{eq:Acc}
\dot{M(t)}_{acc} = \left(\frac{R_{a}}{2a(t)}\right)^2 \dot{M}_{w},
\end{equation}
where $R_{a}$ is the Bondi-Hoyle accretion radius defined as:
\begin{equation}\label{Bondi}
    R_{a} = \frac{2GM_{2}}{{v_{r}}^{2} + {c_{w}}^{2}}
\end{equation}
and where $c_{w}$ is the sound speed of the wind and $v_r$ calculated from the equation \ref{eq:v_r}.
Due to viscous diffusion in the accretion disk, a fraction of the disk mass accretes onto the star itself from the innermost region. At the same time, some of the material can be lost through the outer part of the disk, beyond the Roche radius of the star. In principle, some of the latter material may collide with incoming wind. Nevertheless, the angular size of the disk relative to the companion wind is small, and hence, we neglect this potentially complicated process and assume for simplicity that any material that viscously evolves beyond the Hill radius is lost from the systems. 

\section{Methods} \label{method} 
\subsection{Disk model}
The evolution of the disk is modeled following the $\alpha$-disk prescription (e.g., \cite{lyn+74}; \cite{pri81}) where $\alpha$ represents a dimensionless parameter describing the efﬁciency of the transport of angular momentum.
The kinematic viscosity is defined by
\begin{equation}
\label{eq:viscosity}
    \nu(t, r) = \alpha \  c_{s} \ H(t, r),
\end{equation}
where $c_{s}$ is the sound speed of the gas and $H(t, r)$ is the vertical scale-height of the disk. The $\alpha$ parameter is taken to have a constant value of $0.01$.
The sound speed $c_{s}$ of the gaseous disk depends on the disk mid-plane temperature $T$:
\begin{equation}
    c_{s}(T)= (T)^{0.5} \sqrt{\frac{\gamma \ k}{\mu \ m_{H}}}
\end{equation}
with $\gamma$ the heat capacity ratio and $\mu$ the mean molecular weight, $k$ the Boltzmann constant and $m_H$ the hydrogen atomic mass. 
The vertical scale height is defined as:
\begin{equation}\label{eq:H}
    H(t,r)= \frac{c{_s}(t,r)}{\Omega(r)}, 
\end{equation}

The disk inner cut-off R$_{in}$ is assumed to be at $0.1$ AU, below which mass is assumed to be lost from the disk (accreted to the star).
The disk outer cut-off changes dynamically, as the binary systems evolve. Following the approach in \cite{2013ApJ...764..169P}, we adopt the prescription for the cut-off of the outer radius of the disk $R_{out}$ to be
\begin{equation}\label{eq:R_out}
    R_{out} = r_{l} \ a(t)
\end{equation}
with $r_{l}$ the radius of the Roche limit, from \cite{egg83}:
\begin{equation}\label{eq:r_l}
    r_l =
    \frac{0.49 q_{\star}^{2/3}}{0.6 q_{\star}^{2/3} + ln(1+q_{\star}^{2/3})}
\end{equation}
with $q_{\star}$ the mass ratio $M_{2} / M_{1}$. 
\newline
$a(t)$ is the binary semi-major axis at time $t$.

As we discuss below, any mass that is viscously transported beyond the inner and outer borders of the disk is assumed to be lost from the system.  

\subsection{The temperature profile}
The thermal structure of the disk is evaluated at each time step of the evolution of the disk, following the approach in \cite{2013ApJ...764..169P}. \\
The initial temperature profile is:
\begin{equation}\label{eq:T1}
     T(0, r) =
   T_{0} \ r^{-\frac{1}{2}},
\end{equation}
where $r$ is the distance from the star and $T_{0}$ is the temperature at $1$ AU from the host star, arising from the irradiation by the accreting star.
When the stellar wind is captured by the companion and forms a disk, the mid-plane temperature $T(t,r)$ is determined by several physical processes, each contributing to the temperature: disk heating occurs through viscous dissipation ($T_{\nu}$), irradiation of the central star ($T_{I}$), and the dissipation of kinetic energy from the incoming accreting wind material  ($T_{W}$):
\begin{equation}\label{eq:T1}
     T{^4}(t,r) =
   T_{\nu}{^4}(t,r) + T_{I}{^4}(t,r) + T_{W}{^4}(t,r)
\end{equation}
The viscous contribution $T_{\nu}(t,r)$ is given by
\begin{equation}\label{eq:T_v}
    T_{\nu}{^4}(t,r) = \frac{27}{64} \frac{\kappa}{\sigma} \ \nu(t, r) \ [{\Sigma(t,r)}]^2 \ [\Omega(r)]^2,
\end{equation}
where $\nu(t,r)$ is the kinematic viscosity, $\sigma$ is the Stefan-Boltzmann's constant, and $\kappa$ is the opacity.
\newline
The contribution from the in-falling material onto the disk is given by
\begin{equation}\label{eq:T_w}
    T_{W}{^4}(t,r) = \frac{G}{2 \pi r^2 R_a} \ M_2 \ {{\dot{M}}_{acc}(t)},
\end{equation}
The contribution from irradiation $T_{I}(t,r)$, is  \citep{chi+97}
\begin{equation}\label{eq:T_w}
    T_{I}{^4}(t,r) = \frac{\Theta(t,r)}{2} \ \left(\frac{R_{\star}}{r}\right)^2 \ T_{\star}^4,
\end{equation}
where the factor $\Theta(t,r)$, given the assumption of a vertically isothermal disk, is: 
\begin{equation}\label{eq:T_w}
    \Theta(t,r) = \frac{4}{3 \pi} \left(\frac{R_{\star}}{r}\right)^3 
    \
    +
    \
    R \pdv*{\left(\frac{H(t,r)}{r}\right)}{r},
\end{equation}
with the height scale $H(t,r)$ calculated from the equation \ref{eq:H}.

In the inner region of the disk, the temperature increases sufficiently to vaporize dust, affecting the opacity (\cite{cha09}). 
The opacity $\kappa$ is given by
\begin{equation}\label{eq:opacity}
     \kappa = {\kappa}_0 \left(\frac{T}{T_e}\right)^n,
\end{equation}
where $n = -14$ in regions with $T > T_e = 1380$ K (\cite{rud+91}; \cite{ste98}), and $n = 0$ for $T < T_e$.
\newline
Finally, the equation \ref{eq:T1} is solved numerically using a Newton-Raphson technique.

In our model, we do not consider the radiative heating of the disk by the primary evolving donor star.
However, as argued in \cite{2013ApJ...764..169P}, if the outer disk radiates efficiently, this extra heating has a negligible impact on the structure of the material in the inner disk.

\subsection{Disk evolution}
Our circumstellar disk is modeled as a thin disk. We assume $H(r,t) \ll r$, where $r$ is the radial distance from the star and $H(r,t)$ is the vertical scale-height calculated in the equation \ref{eq:H}.
The disk rotation profile is Keplerian with an orbital angular velocity $\Omega(r)$. 
All the disk quantities are assumed to be dependent only on the radial distance $r$ and time $t$.
Under these assumptions and following the standard approach for constructing a thin-disk model (\cite{1973A&A....24..337S}), the surface density evolution equation $\Sigma(t, r)$ for a circumstellar disk in the presence of a wind accretion is:
\begin{equation}
\label{eq:diffusion}
      \begin{aligned}
        \pdv{\Sigma(t,r)}{t} =
        3 r^{-1}
        \pdv*{\left(r^{1/2}\pdv*{[r^{1/2} \ \nu(t,r) \ \Sigma(t,r)]}{r}\right)}{r} +
        \left(\pdv{\Sigma(t,r)}{t}\right)_{Wind}.
   \end{aligned}
\end{equation}
The first term is the viscous evolution of the surface density solution $\Sigma(t, r)$ and the second one represents its evolution due to the wind accretion.

For initial numerical stability, we initiate the disk with some fiducial low mass of M$_{d,0}$ = 10$^{-8}$ M$_{\odot}$ (this does not affect the results).
\newline
The accreting mass from the wind material, falling into the disk, is given by 
\begin{equation}
 \label{eq:mass_accretion}
    \dot{\Sigma}(t,r) dr =
    \frac{\dot{M(t)}_{acc}}{2 \pi r R_{a}}
    dr,
\end{equation}
where $\dot{M(t)}_{acc}$ is  calculated from  equation \ref{eq:Acc}.
\newline
Following the approach in \cite{bat+81}, we change the variable to $X(r)= 2 r^{1/2}$ and $S(t, X) = X(r)$ $\Sigma(t, r)$ obtaining:
\begin{equation}
    \label{eqn:finite difference}
     \pdv{S}{t} =
     \frac{12}{X^{2}}
     \frac{\partial^2}{\partial X^2}\left(S \nu \right) +
     \frac{4}{X}
     \frac{\dot{M}_{acc}}{2 \pi R_{a}}
\end{equation}

We integrate the equation (\ref{eqn:finite difference}) using a finite difference method implicit in time on a grid with an equal spacing $\Delta r$ in the spatial variable $r$ and a constant time-step $\Delta t$.
The initial surface density profile is taken from the similarity solution of the diffusion equation (\cite{lyn+74}, \cite{har+98}):
\begin{equation}
    \Sigma (0, r) =
    \frac{M_{d,0}}{2 \pi r R_{out}}
    e^{- r / R_{out}},
\end{equation}
where $M_{d,0}$ is the initial disk mass and $R_{out}$ the initial disk radius.

To summarize, the procedure done at each time-step $\Delta t$ is as follows. We compute the temperature radial profile $T(t,r)$ (equation \ref{eq:T1}), then we calculate the viscosity radial profile $\nu (t,r)$ (equation \ref{eq:viscosity}), and, finally, we find the surface radial density profile $\Sigma (t,r)$ (equation \ref{eqn:finite difference}).
%
\begin{figure*}
    \centering
    \includegraphics[width=0.9\textwidth]{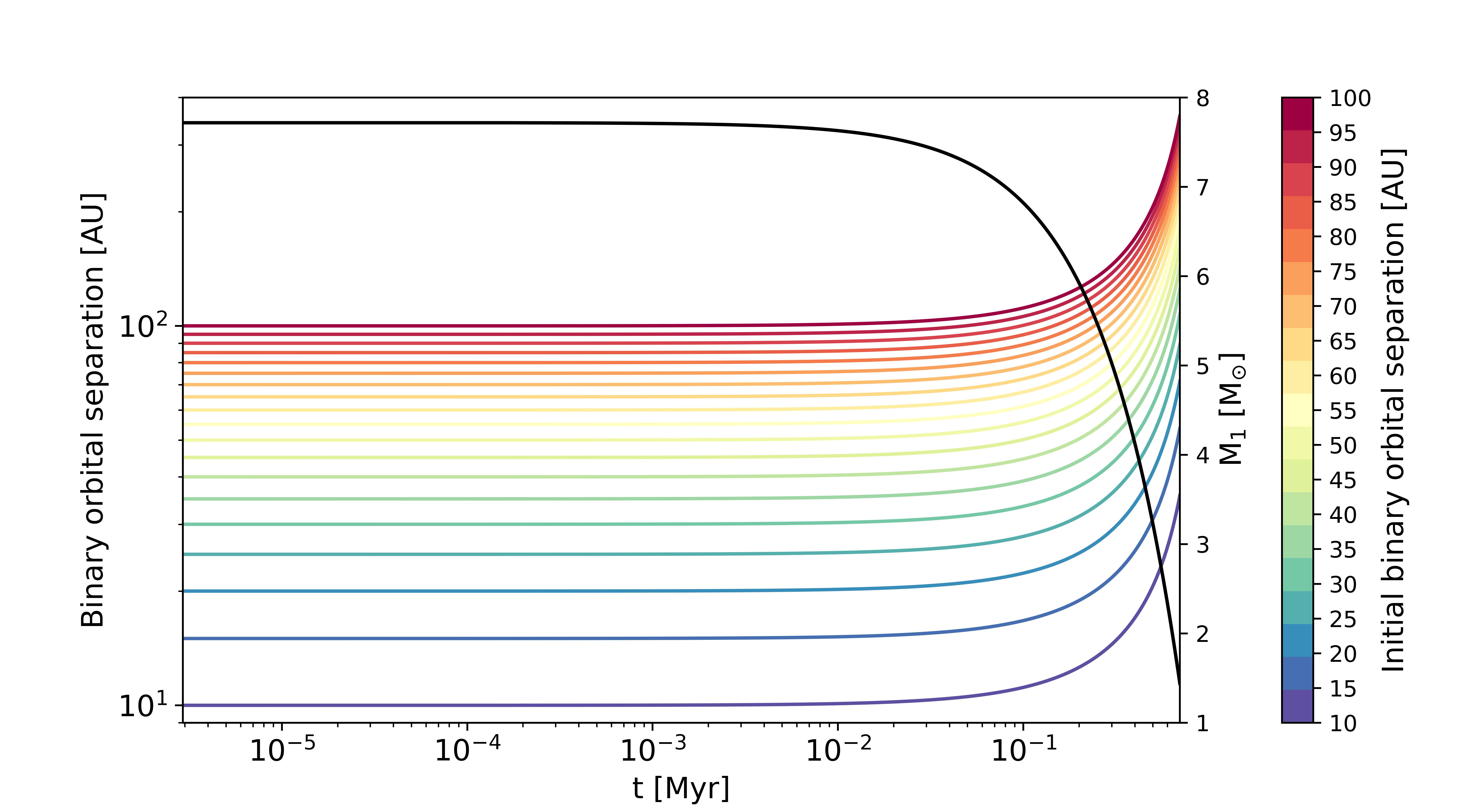}
    \caption{The time evolution of the orbital separation $a(t)$ of the binary system for  model IV.
    Each line corresponds to an initial orbital separation following the color bar on the right.
    The evolution of the mass-losing stellar donor $M_1$ (initially 8 $M_{\odot}$, for this model) is over-plotted (solid black line).
    }
    \label{fig:binary_plots_8M}
\end{figure*}
%
\subsection{Evolution of the binary system}
In our simulations, we take into account the binary orbital expansion due to the mass loss from the donor star.
We assume that the mass loss is a sufficiently slow process with respect to the orbital timescales, and the effect on the orbit can therefore be taken to be adiabatic, in which case the eccentricity can be assumed to be approximately constant.
Following \cite{lag+06}, the adiabatic evolution of the orbital separation of the binary system with total mass $M(t)$, is given by
\begin{equation}\label{eq:T}
     a(t) \ M(t) = constant,
\end{equation}
Therefore, if $M(t)$ decreases, the orbit gets wider. 
In particular, the system expands to a larger final separation:
\begin{equation}\label{eq:T}
     a_f = \frac{m_i}{m_f} a_i,
\end{equation}
where $m_f$ is the final total mass of the system after its evolution, and $m_i$ and $a_i$ are the initial total mass and the initial orbital separation of the binary system.
If the total change of $m_f$ is known, it is possible to derive the final semi-major axis $a_f$.
At each time step, as shown in fig. \ref{fig:binary_plots_8M}, we compute the evolution of the orbital separation $a(t)$ and we substitute this value in the equations \ref{eq:v_s} and \ref{eq:Acc}.

In order to model the evolution, we introduce the accreted material into the diffusion equation \ref{eq:diffusion}, with infalling material only inside the Bondi-Hoyle radius R$_a$, distributed according to the equation \ref{eq:mass_accretion}. 
For each model, we begin the modeling at the beginning of the AGB phase (first thermal pulse), when mass-loss becomes large, and finish when the donor becomes a white dwarf, and no further mass-loss from the donor occurs.
We follow the surface density radial profile $\Sigma(t,r)$ and the temperature radial profile $T(t,r)$.

\subsection{Models grid}

In our study, each binary model we consider is defined by the mass of the progenitor of the donor star $M_1$, and the binary separation (see table \ref{tab:models}). 
We considered four donor masses, 1, 3, 5, and 8 $M_{\odot}$ and a grid of initial orbital separations $a(0) = a_0$ in the range $10$ AU to $100$ AU, with a size-step of $5$ AU. 
For simplicity, we adopt a constant mass $M_2 = 1 M_{\odot}$ for the accretor, main sequence star. 
We set v$_{w} = 10$ km/s and $c_{w} = 20$ km/s respectively in the equation \ref{eq:v_r} and \ref{Bondi}. We consider the evolution of the systems from the first AGB stage, through the second AGB stage, until the end of the AGB and the formation of the final WD. 
Table \ref{tab:models} summarizes the model parameters, the timescales for each stage, and the primary (donor) masses at these stages.

\section{Results}
\begin{figure}
    \centering
    \includegraphics[width=0.49\textwidth]{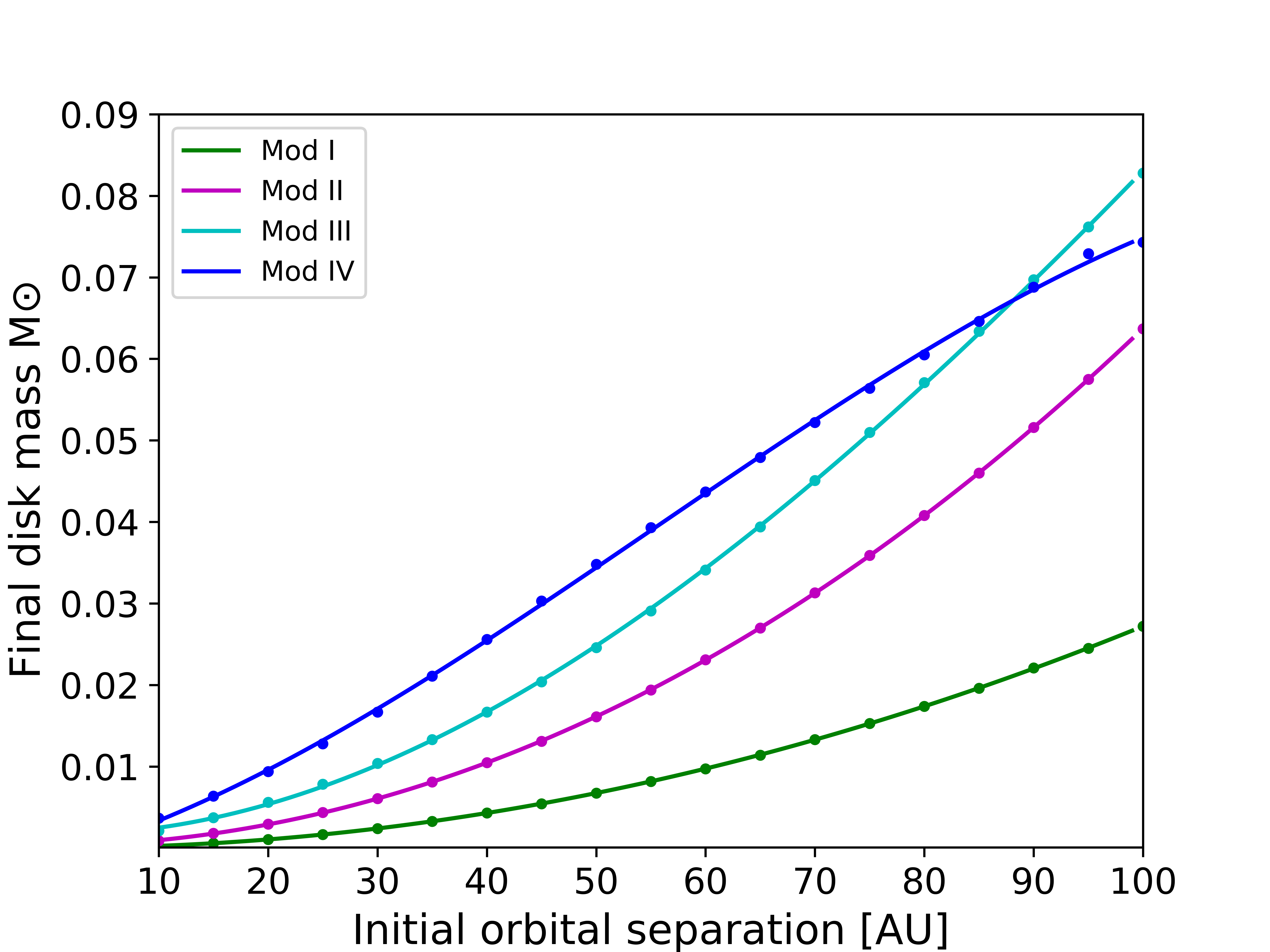}
    \caption{Final mass of the disk in $M_{\odot}$ as a function of the initial orbital separation of the binary system $a_0$ in AU. 
    Each line represents a model. 
    }
    \label{fig:disk_mass}
\end{figure}

Figures \ref{fig:binary_plots_8M}-\ref{fig:disk_mass} and Tables \ref{table:results1} and \ref{table:results2} summarize our main results. The evolution of the binary separation is shown in Fig. \ref{fig:binary_plots_8M}, and the evolution of the disk density and temperature profiles
can be seen in  Figs. \ref{fig:sigma_temperature_1M}-\ref{fig:sigma_temperature_8M}. The evolution of the disk mass and accretion rate onto the accretor star are shown in Fig. \ref{fig:disk_mass}.

As can be seen in Figs.\ref{fig:sigma_temperature_1M}-\ref{fig:sigma_temperature_8M}, the disks evolve in time, 
and reach a quasi-stable configuration and an approximately steady accretion rate onto the central star (Fig.\ref{fig:disk_mass}, in timescales shorter than the AGB lifetime of the donor. This is generally consistent with the results obtained by PK13. Note, however, that given that, unlike the PK13 study, we model the realistic binary expansion and non-constant mass-loss, we never reach a truly steady state, but only a quasi-steady state.

As can be seen in Table \ref{table:results1} and Fig. \ref{fig:disk_mass} the final masses of the modeled disks, can be significant. In particular, the disk masses  exceed the Minimum Solar Mass Nebulae $\sim 10^{-2} M_{\odot}$, for binaries with wide separations.

The accretion rates $\dot{M}_{in}$ onto the star $M_{2}$ show a common general trend in all the models: it decreases with the increase of the orbital separations of the binary systems (for more details see results in tables \ref{table:results1} and \ref{table:results2}). Therefore although more compact systems directly capture more mass, and show higher accretion onto the star itself, their disks are cut at smaller outer radii and overall hold less (total) mass as they reach a quasi-steady state.
For these systems, the increase of the disk mass is correlated with the diffusive expansion of the outer disk radius $R_{out}$: in massive models ($M_1^{in} > 3M_{\odot}$), the final disk radius increases by more than $50\%$ of its initial value (see column $8$ and $9$ in the tables of results \ref{table:results1} and \ref{table:results2}).
Moreover, for the cases with wide orbital separations, the relative velocity $v_r$ decreases, consequently the Bondi-Hoyle radius R$_a$ increases, allowing for the formation of more massive disks.

\begin{deluxetable*}{l|c|c|c|c|c|c|c|c|c|c|l}
\label{table:models}
\tablewidth{0pt} 
\tablecaption{Properties of the simulated models. \label{tab:models}}
\tablehead{
\colhead{Model} &  
\colhead{M$_{accretor}$} & 
\colhead{M$_{donor}$} & 
\colhead{1$^{st}$AGB} & 
\colhead{1$^{st}$ AGB } & 
\colhead{2$^{nd}$AGB } & 
\colhead{2$^{nd}$ AGB } & 
\colhead{WD} & 
\colhead{WD} & 
\colhead{Total time} & 
\colhead{Total mass loss} & 
\colhead{Separation range} \\
\colhead{} &  
\colhead{ (M$_{\odot}$)} & 
\colhead{ (M$_{\odot}$)} & 
\colhead{ (Myr)} & 
\colhead{ (M$_{\odot}$)} & 
\colhead{ (Myr)} & 
\colhead{ (M$_{\odot}$)} & 
\colhead{ (Myr)} & 
\colhead{ (M$_{\odot}$)} & 
\colhead{ (Myr)} & 
\colhead{ (M$_{\odot}$)} & 
\colhead{ (AU)}
} 
\startdata 
I & $1$ & $1$ & $1245.9$ & $0.73$ & $1246.3$ & $0.59$ & $1246.4$ & $0.52$  & $4.8$  & $0.21$ & $10-100$ \\
\hline 
II & $1$ & $3$ & $473.4$ & $2.97$ & $476.4$ & $2.95$ & $477.6$ & $0.75$ & $4.2$ & $2.22$ & $10-100$ \\
\hline 
III & $1$ & $5$ & $120.1$ & $4.93$ & $120.9$ & $4.89$ & $121.5$ & $0.99$ & $1.4$ & $3.94$ & $10-100$ \\
\hline 
IV & $1$ & $8$ & $41.8$ & $7.72$ & $42.0$ & $7.67$ & $42.5$ & $1.44$ & $0.7$ & $6.28$ & $10-100$ \\
\enddata
\end{deluxetable*}
\begin{deluxetable}{l|c|c|c|l}
\tablecaption{Table of coefficients of the polynomial fit in the equation \ref{eq:equation_model_IV}. \label{tab:coefficients}}
\tablehead{
\colhead{Model} & 
\colhead{C$_{1}$} & 
\colhead{C$_{2}$} & 
\colhead{C$_{3}$} & 
\colhead{C$_{4}$}
} 
\startdata 
I   & $-2.15\times10^{-10}$ & $2.78\times10^{-6}$ & $-4.29\times10^{-6}$ & $3.72\times10^{-5}$ \\
II  & $ 2.12\times10^{-9}$ & $6.02\times10^{-6}$ & $ 1.15\times10^{-5}$ & $2.49\times10^{-4}$ \\
III & $-2.76\times10^{-8}$ & $1.12\times10^{-5}$ & $-2.69\times10^{-5}$ & $1.68\times10^{-3}$ \\
IV  & $-5.87\times10^{-8}$ & $9.78\times10^{-6}$ & $ 3.71\times10^{-4}$ & $-1.26\times10^{-3}$ \\
\enddata
\end{deluxetable}
\subsection{Disk structure}
\begin{figure*}
    \centering
    \includegraphics[width=0.9\textwidth]{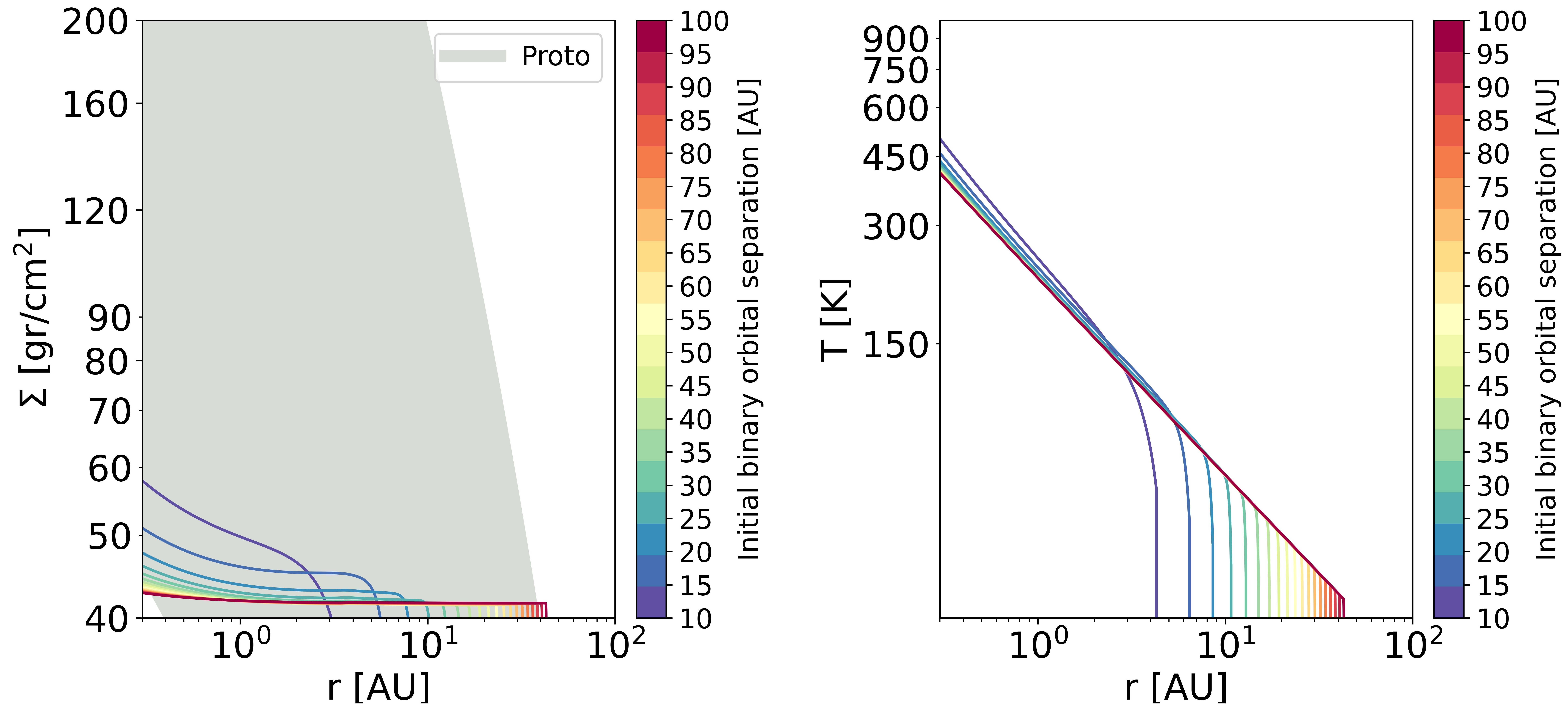}
    \caption{Model I: 1 M$_{\odot}$ donor.
    The final radial surface density profile of the accretion disk, $\Sigma(t_{fin},r)$. Each line corresponds to an initial orbital separation shown in the color bar. The shaded region represents the range of protoplanetary disk models inferred from observations (\cite{Andrews_2010}).
    }
    \label{fig:sigma_temperature_1M}
\end{figure*}

\begin{figure*}
    \centering
    \includegraphics[width=0.9\textwidth]{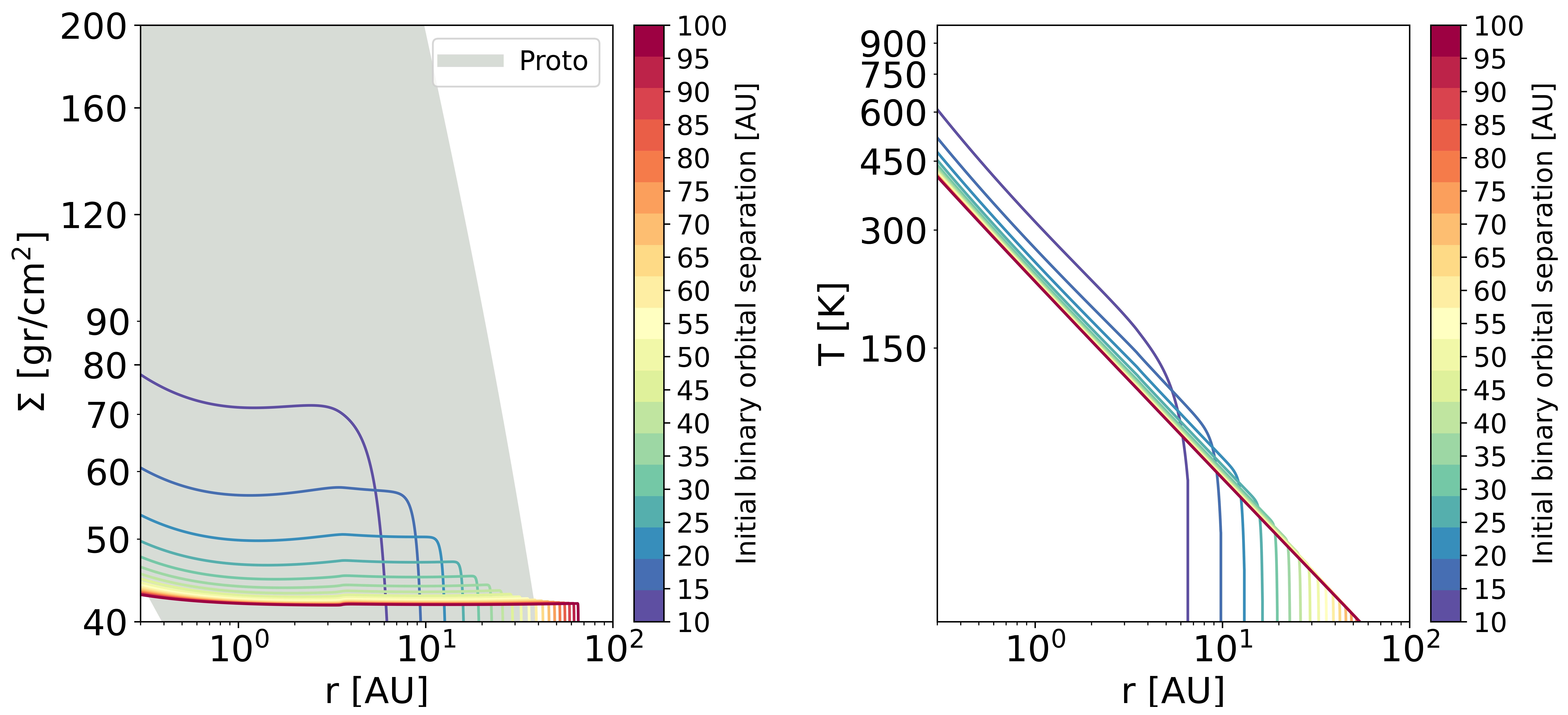}
    \caption{Model II: .
    Similar to Fig. \ref{fig:sigma_temperature_1M} but for the 3 M$_{\odot}$.
    }
    \label{fig:sigma_temperature_3M}
\end{figure*}

\begin{figure*}
    \centering
    \includegraphics[width=0.9\textwidth]{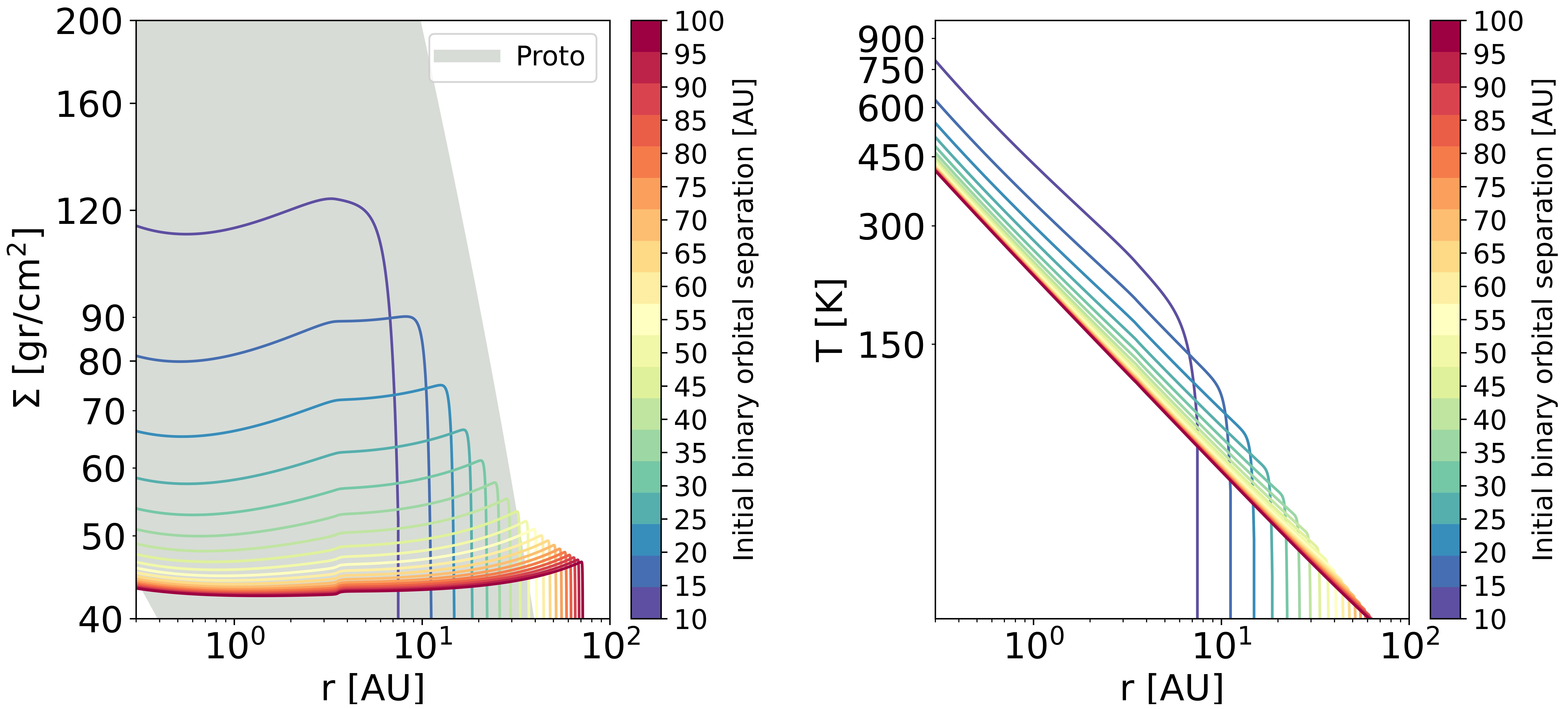}
    \caption{Model III: 5 M$_{\odot}$ donor.
     Similar to Fig. \ref{fig:sigma_temperature_1M} but for the 5 M$_{\odot}$.
    }
    \label{fig:sigma_temperature_5M}
\end{figure*}

\begin{figure*}
    \centering
    \includegraphics[width=0.9\textwidth]{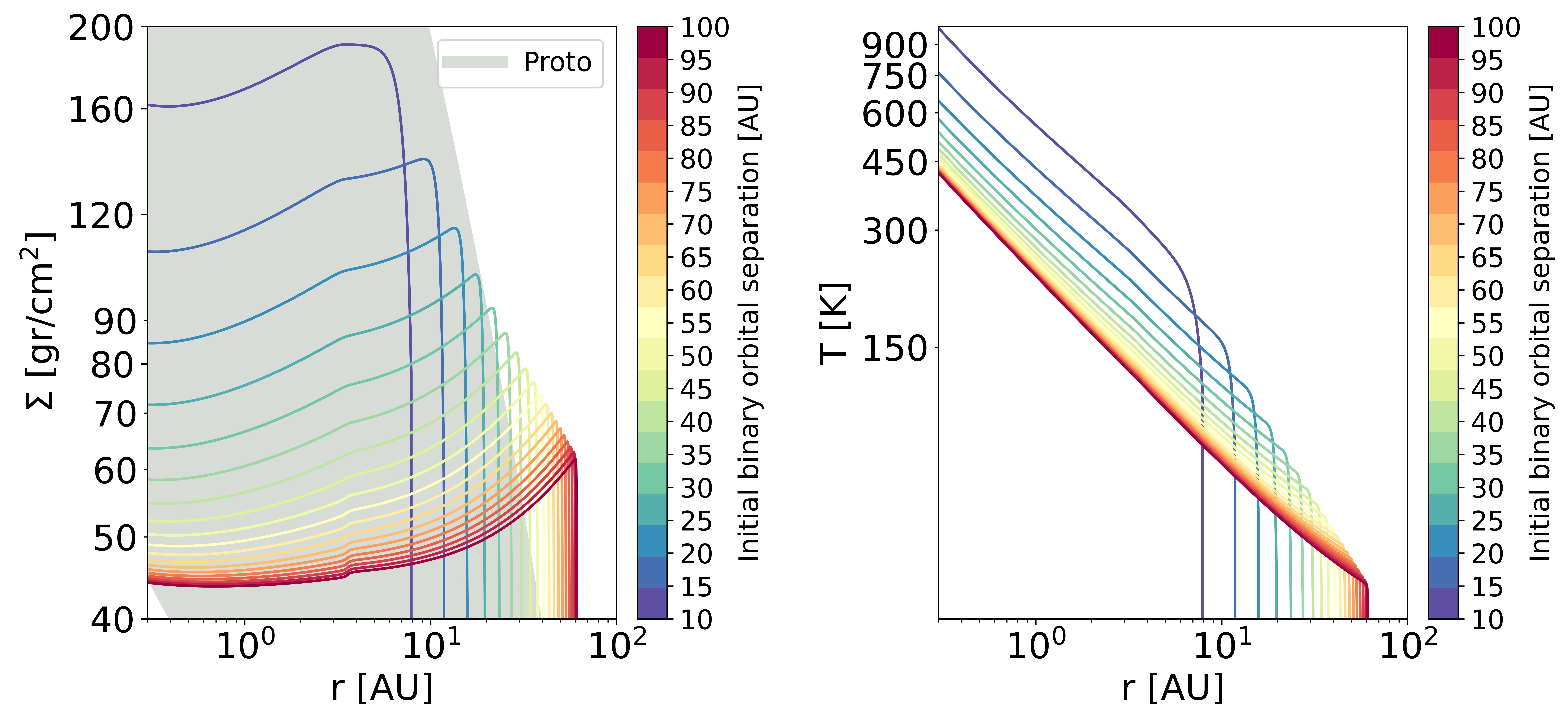}
    \caption{Model IV: 8 M$_{\odot}$ donor.
     Similar to Fig. \ref{fig:sigma_temperature_1M} but for the 8 M$_{\odot}$.
    }
    \label{fig:sigma_temperature_8M}
\end{figure*}

At the end of the donor AGB phase, the final surface density profile $\Sigma(t_{fin},r)$, in most cases, overlaps with the typical density range of observed protoplanetary disks (see figures \ref{fig:sigma_temperature_1M}-\ref{fig:sigma_temperature_8M}; greyed regions show the typical regions of protoplanetary disks density profiles ).
For each model, we show the correlation between the final mass of the disk M$_{disk}$ and the initial orbital separation of the binary system a$_0$ (see figure \ref{fig:disk_mass}). 
These relations, although restricted to a limited number of models, represent a useful tool to calculate the initial conditions for studying the evolution of disks after the donor AGB phase.

\begin{figure*}
    \centering
    \includegraphics[width=0.75\textwidth]{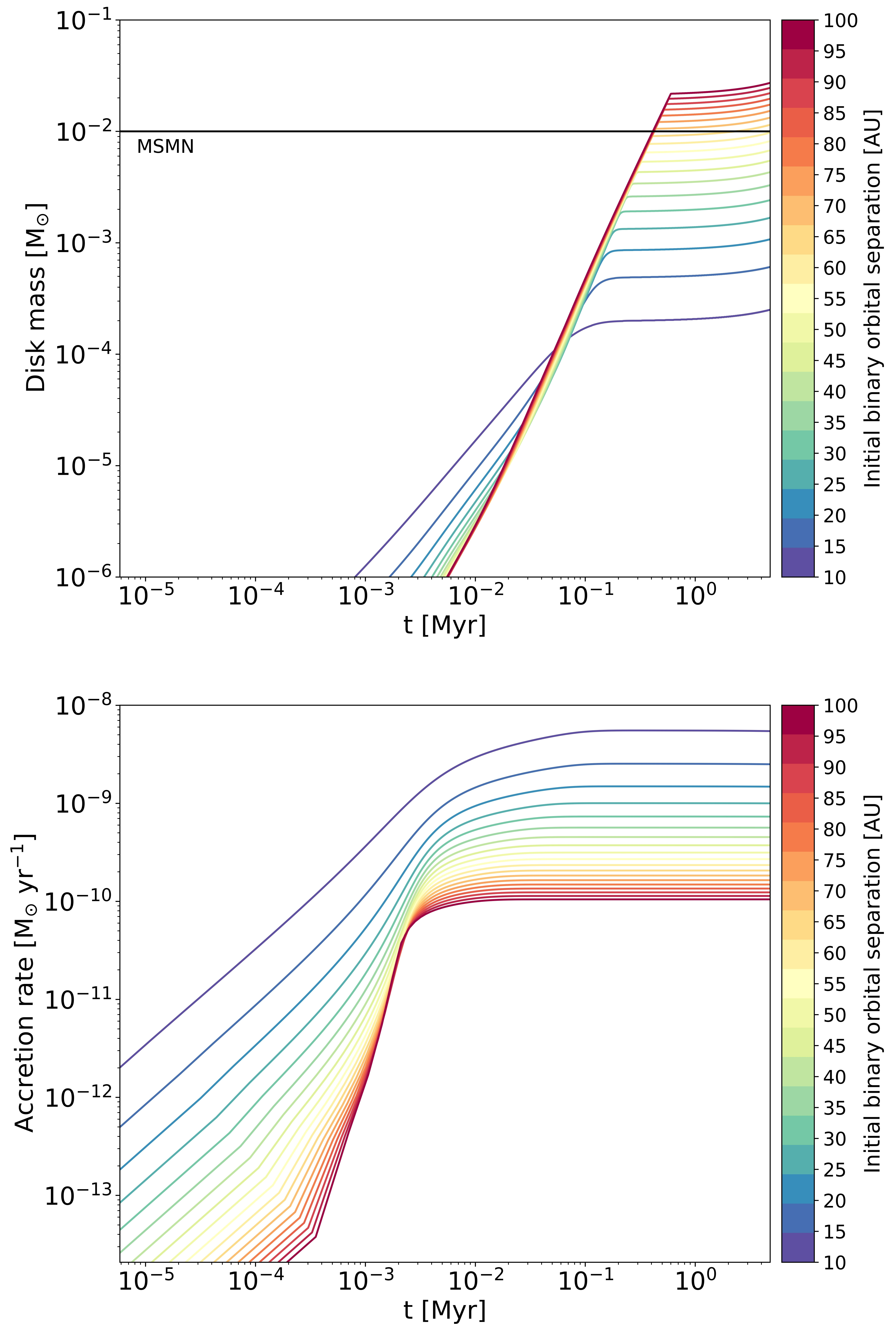}
    \caption{Model I: 1 M$_{\odot}$
    Top panel: The time evolution of the disk mass.
    The black solid line represents the Minimum Solar Mass Nebulae (MSMN) $\sim 10^{-2} M_{\odot}$.
    Each line corresponds to a different initial orbital separation shown in the color bar.
    Bottom panel: The time evolution of the accretion rate onto the main sequence star.}
    \label{fig:disk_plots_1M}
\end{figure*}

One can fit (polynomial fit) the correlation between  the final mass of the disk M$_{disk}$ and the initial orbital separation of the system a$_0$, for the different models:
\begin{equation}
 \label{eq:equation_model_IV}
    M_{disk} = C_1 \ a_0^3 + C_2 \ a_0^2 + C_3 \ a_0 + C_4
\end{equation}
In model I, for initial orbital separations of $a_0 > 55$, we find the disks develop an approximately flat slope for the final surface density profile (see the first panel in figure \ref{fig:sigma_temperature_1M}).
The final temperature profile of the mid-plane temperature follows a broken power law, with more compact binary systems reaching higher temperatures in the inner region of the disk (see the panel on the right in fig. \ref{fig:sigma_temperature_1M}).
Finally, the mass accretion rates onto the central star show higher values for close binary systems, roughly a factor $\sim 2$ larger than the wider ones with $a>45$ (see the bottom panel in figure \ref{fig:disk_plots_1M}). 
\newpage
%
%
%
%
\begin{deluxetable}{llll}
\tabletypesize{\footnotesize}
\tablewidth{0pt} 
\tablecaption{Table of fixed parameters used in our simulations. \label{tab:parameters}}
\tablehead{
\colhead{Parameter} & 
\colhead{Value} & 
\colhead{Unit} & 
\colhead{Description} 
} 
\startdata 
{$R_{in}$}   & $0.1$                 & AU                                       & Inner radius of the disk \\
{$R_{sun}$}  & $4.65 \times10^{-3}$  & AU                                       & Radius star $M_2$ \\
{$T_{star}$} & $5.78\times10^{3}$    & K                                        & Temperature star $M_2$ \\
{$M_d$}      & $10^{-8}$             & $M_{\odot}$                              & Initial mass of the disk \\
{$\alpha$}   & $10^{-2}$             & -                                        & $\alpha$ disk parameter \\ 
{$\kappa$}   & $8.94\times10^{-16}$  & $M_{\odot} \ yr^{-3} \ K^{-4}$           & Stefan-Boltzmann constant \\
{$\sigma$}   & $3.08\times10^{-61}$  & $AU^2 \ M_{\odot} \ yr^{-2} \ K^{-1}$    & Boltzmann constant \\
{$\gamma$}   & $1.66$                & -                                        & Ratio specific heats \\  
{$\mu$}      & $8.39\times10^{-58}$  & $M_{\odot}$                              & Mean molecular weight \\
{$k$}        & $2.67\times10^{7}$    & $AU^2 \ M_{\odot}^{-1}$                  & Opacity
\enddata
\end{deluxetable}
%
In model II, the final surface density profile $\Sigma(t_{fin},r)$ shows a bump in the inner region of the disk, forming at the Bondi-Hoyle accretion radius R$_a$ (see the left panel in figure \ref{fig:sigma_temperature_3M}). 
Beyond this radius, it shows an approximately flat shape, in particular for systems with initial orbital separations wider than $30$ AU.
This peculiar \textit{bump} represents the final result of an efficient diffusion mechanism of the gaseous mass from the hot and massive inner regions to the cooler and less massive regions of the disk, beyond the R$_a$ radius.
%
%
%
In model III, in systems with an initial orbital separation $a_0 < 40$ AU, higher temperatures, with respect to the previous model I and II, are reached close to the central star. We also note that higher temperatures in the inner region of the disk (see the bottom panel in figure \ref{fig:sigma_temperature_5M}) increase the inward diffusion of gas, and, consequently, the slopes of the final surface density profiles are smoother with respect to the previous models I and II.
In model IV, the final surface density profiles $\Sigma(t_{fin},r)$ (see the top panel of the figure \ref{fig:sigma_temperature_8M}) show an interesting and peculiar shape: the radial surface density increases at large radii. 
%
%
%
In particular, inside the Bondi-Hoyle radius, the mid-plane temperature increases, due to the significant heating from from the infalling material (parameterized by $T_{W}$; see eq. \ref{eq:T_w}) and by the viscous heating $T_{\nu}$ (see eq. \ref{eq:T_v}).
In the inner regions of the disk, higher temperatures allow a more efficient viscous diffusion, both the inward and the outward direction (see the temperature profiles in the bottom panel of the figure \ref{fig:sigma_temperature_8M}). 
Consequently, we experience a large mass accretion onto the central star, in the inward direction, and a large spreading out of gas, in the outward direction.
Similarly, due to the decrease in temperature in the outer regions, viscous diffusion is less efficient.
As we can see in figure \ref{fig:disk_mass}, this model does not follow the general trend of the previous models, in particular for $a_0 > 70$ AU.

\section{Discussion}
\label{sec:discussion}
\subsection{The case of Mira}
Mira (Omicron Ceti) is a binary stellar system, consisting of a red giant donor star, Mira A, and a confirmed white dwarf companion, Mira B (\cite{ire+07}), which shows evidence of a circumstellar disk, likely formed from accreted mass from the companion wind. As such, the Mira system provides an interesting test case for the formation of second-generation disks. We therefore explore a model of this system and compare our results with the current observations.

The donor in this system has a mass of $1.18$ M$_{\odot}$ with a measured mass loss rate of $ 4.40 \times 10^{-7}$ $M_{\odot}$ $yr^{-1}$ (\cite{ire+07}).
The accretor star Mira B is a white dwarf with a mass of $0.6$ $M_{\odot}$ and an accretion rate of $ \sim 10^{-10}$ $M_{\odot}$ $yr^{-1}$ (\cite{sok+10}).
The current orbital separation of the system is $ \sim 75$ AU (\cite{ire+07}). 
We simulated the dynamical evolution of the Mira system for an evolution time of $1$ Myr, using the methodology presented in the section \ref{method}. We assume a constant inner radius for the disk of $0.1$ AU around the WD accretor, motivated by the possibility of the erosion of the inner region of the disk, due to the stellar magnetic field or eruptive phenomenons in the photo-sphere of this white dwarf.

\begin{figure*}
    \centering
    \includegraphics[width=\textwidth]{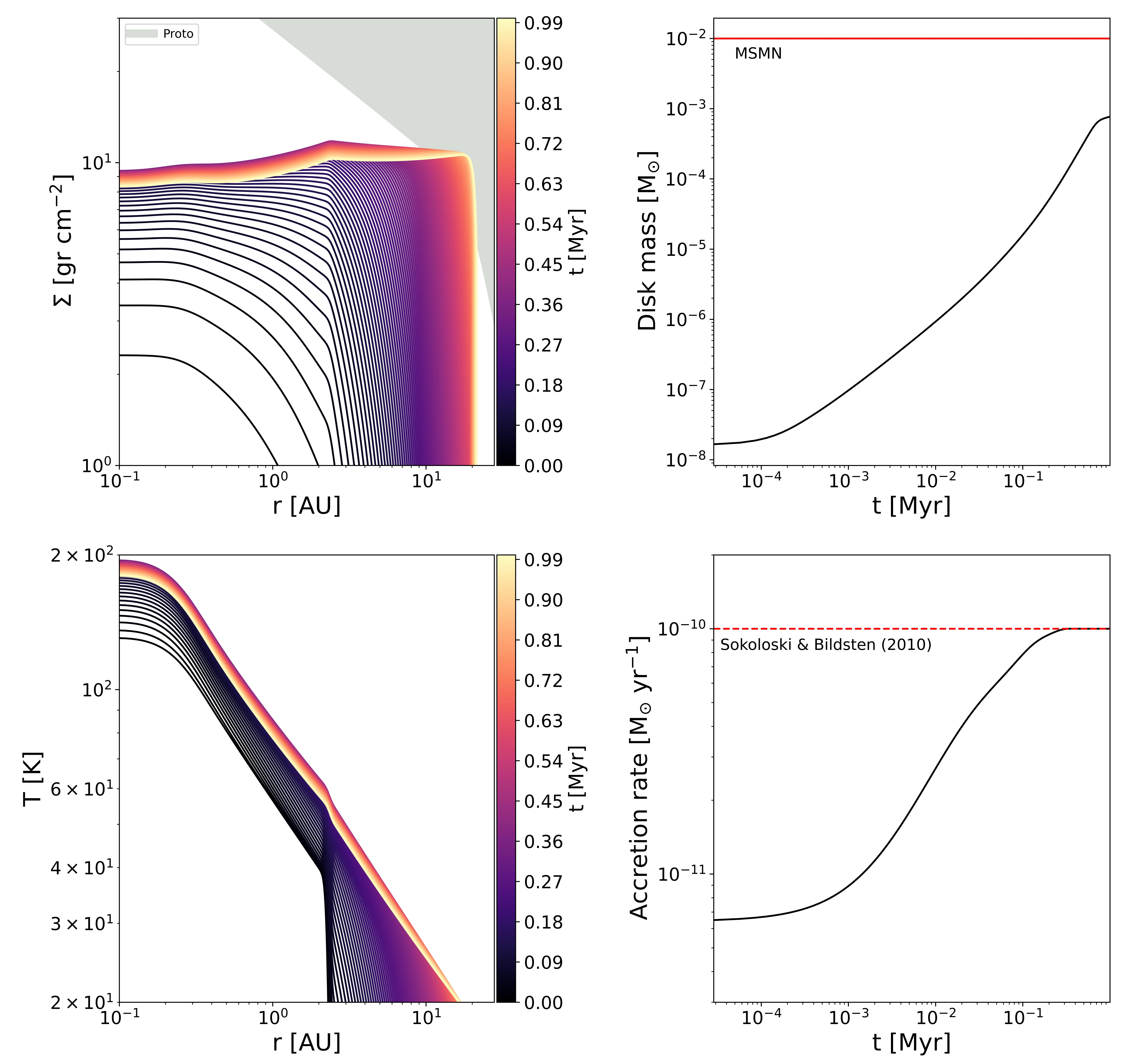}
    \caption{Mira system. The top left panel shows the radial surface density profile $\Sigma(t,r)$.
    Each line corresponds to a time step in the color bar. The shaded region represents the range of protoplanetary disk models inferred from observations (\cite{Andrews_2010}).
    The bottom left panel shows the radial temperature profile $T(t,r)$. 
    The top right panel shows the time evolution of the disk mass. A constant Minimum Solar Mass Nebulae (MSMN) $10^{-2}$ $M_{\odot}$ is over-plotted in a red solid line.
    The bottom right panel shows the time evolution of the mass accretion rate onto the white dwarf, solid black line, with the observed value, dashed red line, proposed by \cite{sok+10}.}
    \label{fig:mira}
\end{figure*}

We define a \textit{steady state} when the change in disk mass per unit time is smaller than $10^{-3}$ of the input rate.
Considering the current mass of the red giant Mira A of $1.18$ M$_{\odot}$ with a constant mean mass loss of $4.40 \times 10^{-7}$ $M_{\odot}$ $yr^{-1}$, we set the mass of the progenitor to be $1.62 M_{\odot}$ with an initial orbital separation of $60$ AU.
As shown in the bottom right panel in figure \ref{fig:mira}, the system reaches a quasi-steady state configuration in $\sim 1$ Myr.

Following the evolution, we find the temperature profile increases till the maximum value $\sim 2 \times 10^{2}$ K at the inner radius (see the bottom left panel in figure \ref{fig:mira}).
After a quasi-steady state is achieved, it gradually cools, following a typical broken power law. 
The final accretion rate onto the star is $1.03 \times 10^{-10}$ $M_{\odot}$ $yr^{-1}$ and the final mass of the disk is $0.82\times10^{-3}$ $M_{\odot}$, in agreement with the values proposed by \cite{sok+10}.
Despite the promising outcomes, the model used to simulate this system suffers from some approximations, some of which are mentioned in section \ref{sec:Model limitations}. 
We neglect potential nova outbursts generated by the infalling material on the surface of the white dwarf, which may have a recurrence time of $\sim 10^{5}$ yr (\cite{sok+10}).
These events might dramatically perturb the inner region of the disk, increasing the size of the inner cutoff and affecting the long-term secular dynamical evolution of the disk.
Future theoretical investigations will design a more realistic model. 

\subsection{Observational implications}
Given our finding of the likely existence of SG protoplanetary disks, we propose to search for such disks, in wide-evolved binary systems, using the same tools used for current observations of protoplanetary disks (e.g. ALMA). These should be found either in cases of wide systems in which one of the components is on the AGB, or potentially in wide systems, where a very young WD exists, i.e. a short time after it finished its AGB stage, and a remnant disk likely still exists. 

Furthermore, since newly formed planets and/or accreting pre-existing planets would be hot, they might be detected through direct imaging, as currently done for young exoplanetary systems. Indeed, the direct detection of planets in old evolved wide binary systems would provide a smoking gun signature for the existence of second-generation planets (or regrowing first-generation planets).

\subsection{Model limitations}
\label{sec:Model limitations}
We summarize here some simplifying assumptions and limitations of our study. 
First of all, we assume our disc is circular and aligned with the binary orbit with a constant zero eccentricity. 
We neglect the erosion from the photo-evaporation from the accretor star (M$_2$) and its stellar evolution.  
We also assume, for simplicity, a constant opacity originating from gas without any dust contribution. 
The $\alpha$ parameter is a constant value in all our simulations and it represents an approximated approach to model the turbulence in viscous gaseous circumstellar structures.
Finally, our simulations are in 1D in the $r$ variable, assuming constant the components in the other dimensions.
\section{Summary}
We have investigated the formation and the dynamical evolution of second-generation circumstellar disks in wide, formed evolved binary systems. Previous studies explore simplified models with constant mass-loss and binary orbits. Here we have used realistic stellar evolution models for the non-constant mass-loss rates from the mass-losing evolved star, and account for the evolving separations of the systems.

We considered four models, characterized by a different mass of the progenitor stellar donor, respectively 1, 3, 5, and 8 M$_{\odot}$.
For each model, we set initial orbital separations in the range $10$ AU - $100$ AU, with a size-step of $5$ AU.
All the models show the formation of stable viscous circumstellar structures during the stellar evolution of the donor companion.
A quasi-steady state is achieved in all the models, in timescales shorter than the stellar evolution timescale of their stellar donor. 
In the model I and II, the $42\%$ and $68\%$ of the systems reach the Minimum Solar Mass Nebula, and, in the model III and IV, respectively, the $79\%$ and $84\%$.
As expected, massive stellar companions form massive second-generation circumstellar disks.
Systems with wide initial orbital separations ($a_0 > 70$) reach the Minimum Solar Mass Nebula in timescales shorter than the lifetime of the red giant phase of their stellar donor.
The final outer radii of disks increase, with respect to their initial size, from $12\%$ in Model I to $72\%$ in Model IV due to the viscous diffusion.
In models III and IV the final radial surface density profiles $\Sigma(t_{fin},r)$ show significant masses in the outer  regions of the disks.
In most cases, the final radial surface density profiles we find, show similar properties to those of observed, normal (first-generation) protoplanetary disks.
This similarity, represents an important outcome of our investigation, supporting the possibility that second-generation disks could serve as fertile ground for second-generation planet formation (\cite{per10a}, \cite{2019MNRAS.487.3069K}) and/or give rise to revived planet-disk interactions of pre-existing (or exchanged; \cite{kra+12}) planets with the newly formed disk (e.g. leading to regrowth and/or migration and potential destabilization in multi-planet systems.

Finally, we investigated the \textit{Mira} wide evolved binary system. Our model predicts the presence of a second-generation disk with a final mass of $0.82\times10^{-3}$ $M_{\odot}$, formed in a timescale of $ \sim 1$ Myr. We find the final accretion rate onto the white dwarf to be $\sim 10^{-10}$. These properties are consistent with the inferred properties of the observed disk and accretion rates \cite{ire+07,sok+10}.

\section*{Acknowledgements}
We acknowledge support for this project from the European Union's Horizon 2020 research and innovation program under grant agreement No 865932-ERC-SNeX.
%
%
%

\begin{deluxetable*}{lcc|cc|c|c|cc|c|cc|c|c}
\label{table:results1}
\tabletypesize{\scriptsize}
\tablewidth{0pt} 
\tablecaption{Table of results model I and model II. \label{tab:results}}
\colnumbers
\tablehead{
\colhead{Model} & 
\colhead{a$^{in}$} & 
\colhead{a$^{fin}$} & 
\colhead{ M$_1^{in}$} & 
\colhead{ M$_1^{fin}$} & 
\colhead{ M$_2$} & 
\colhead{ R$_a$} &
\colhead{ R$_{out}^{in}$} &
\colhead{ R$_{out}^{fin}$} &
\colhead{ $\dot{M}_{loss}$} & 
\colhead{ $\dot{M}_{disk}^{in}$} & 
\colhead{ $\dot{M}_{disk}^{fin}$} & 
\colhead{ $M_{disk}$} & 
\colhead{$\dot{M}_{in}$}
} 
\startdata 
{       } &10 &11.36  &1 &0.52 &1 &3.259 &3.78  &4.31  & $4.30\times10^{-8}$ & $1.142\times10^{-9}$ & $8.85\times10^{-10}$ & $2.51\times10^{-4}$ &$5.45\times10^{-9}$ \\ 
{       } &15 &17.04  &1 &0.52 &1 &3.350 &5.68  &6.45  &$4.30\times10^{-8}$ & $5.36\times10^{-10}$ & $4.16\times10^{-10}$ & $6.07\times10^{-4}$ &$2.50\times10^{-9}$ \\ 
{       } &20 &22.72  &1 &0.52 &1 &3.398 &7.57  &8.60  &$4.30\times10^{-8}$ & $3.10\times10^{-10}$ & $2.40\times10^{-10}$ & $1.07\times10^{-3}$ &$1.48\times10^{-9}$ \\ 
{       } &25 &28.40  &1 &0.52 &1 &3.427 &9.47  &10.76 &$4.30\times10^{-8}$ & $2.02\times10^{-10}$ & $1.57\times10^{-10}$ & $1.68\times10^{-3}$ &$9.98\times10^{-10}$ \\  
{       } &30 &34.08  &1 &0.52 &1 &3.447 &11.36 &12.91 &$4.30\times10^{-8}$ & $1.42\times10^{-10}$ & $1.09\times10^{-10}$ & $2.41\times10^{-3}$ &$7.31\times10^{-10}$ \\ 
{       } &35 &39.76  &1 &0.52 &1 &3.461 &13.26 &15.06 &$4.30\times10^{-8}$ & $1.05\times10^{-10}$ & $8.14\times10^{-11}$ & $3.29\times10^{-3}$ &$5.64\times10^{-10}$ \\ 
{       } &40 &45.44  &1 &0.52 &1 &3.472 &15.15 &17.21 &$4.30\times10^{-8}$ & $8.09\times10^{-11}$ & $6.27\times10^{-11}$ & $4.30\times10^{-3}$ &$4.52\times10^{-10}$ \\ 
{       } &45 &51.12  &1 &0.52 &1 &3.480 &17.05 &19.36 &$4.30\times10^{-8}$ & $6.43\times10^{-11}$ & $4.98\times10^{-11}$ & $5.45\times10^{-3}$ &$3.73\times10^{-10}$ \\  
{Model I} &50 &56.79  &1 &0.52 &1 &3.487 &18.94 &21.52 &$4.30\times10^{-8}$ & $5.23\times10^{-11}$ & $4.05\times10^{-11}$ & $6.75\times10^{-3}$ &$3.14\times10^{-10}$ \\ 
{       } &55 &62.48  &1 &0.52 &1 &3.492 &20.84 &23.67 &$4.30\times10^{-8}$ & $4.33\times10^{-11}$ & $3.36\times10^{-11}$ & $8.17\times10^{-3}$ &$2.69\times10^{-10}$ \\ 
{       } &60 &68.15  &1 &0.52 &1 &3.497 &22.73 &25.82 &$4.30\times10^{-8}$ & $3.65\times10^{-11}$ & $2.83\times10^{-11}$ & $9.74\times10^{-3}$ &$2.34\times10^{-10}$ \\ 
{       } &65 &73.83  &1 &0.52 &1 &3.501 &24.62 &27.97 &$4.30\times10^{-8}$ & $3.12\times10^{-11}$ & $2.42\times10^{-11}$ & $1.14\times10^{-2}$ &$2.06\times10^{-10}$ \\ 
{       } &70 &79.51  &1 &0.52 &1 &3.504 &26.52 &30.12 &$4.30\times10^{-8}$ & $2.69\times10^{-11}$ & $2.09\times10^{-11}$ & $1.33\times10^{-2}$ &$1.83\times10^{-10}$ \\ 
{       } &75 &85.19  &1 &0.52 &1 &3.507 &28.41 &32.28 &$4.30\times10^{-8}$ & $2.35\times10^{-11}$ & $1.82\times10^{-11}$ & $1.53\times10^{-2}$ &$1.64\times10^{-10}$ \\ 
{       } &80 &90.87  &1 &0.52 &1 &3.509 &30.31 &34.43 &$4.30\times10^{-8}$ & $2.07\times10^{-11}$ & $1.60\times10^{-11}$ & $1.74\times10^{-2}$ &$1.48\times10^{-10}$ \\ 
{       } &85 &96.55  &1 &0.52 &1 &3.512 &32.20 &36.58 &$4.30\times10^{-8}$ & $1.83\times10^{-11}$ & $1.42\times10^{-11}$ & $1.96\times10^{-2}$ &$1.35\times10^{-10}$ \\ 
{       } &90 &102.23 &1 &0.52 &1 &3.514 &34.10 &38.73 &$4.30\times10^{-8}$ & $1.64\times10^{-11}$ & $1.27\times10^{-11}$ & $2.21\times10^{-2}$ &$1.23\times10^{-10}$ \\ 
{       } &95 &107.91 &1 &0.52 &1 &3.515 &35.99 &40.88 &$4.30\times10^{-8}$ & $1.47\times10^{-11}$ & $1.14\times10^{-11}$ & $2.45\times10^{-2}$ &$1.13\times10^{-10}$ \\ 
{       } &100&113.59 &1 &0.52 &1 &3.517 &37.89 &43.04 &$4.30\times10^{-8}$ & $1.33\times10^{-11}$ & $1.03\times10^{-11}$ & $2.72\times10^{-2}$ &$1.05\times10^{-10}$ \\
\hline
{        } &10 &22.69   &3 &0.75 &1 &2.536 &2.88  &6.55  &$5.29\times10^{-7}$ & $8.51\times10^{-9}$  & $1.65\times10^{-9}$  & $9.13\times10^{-4}$ &$1.01\times10^{-8}$ \\ 
{        } &15 &34.03   &3 &0.75 &1 &2.803 &4.33  &9.83  &$5.29\times10^{-7}$ & $4.62\times10^{-9}$  & $8.97\times10^{-10}$ & $1.81\times10^{-3}$ &$4.17\times10^{-9}$ \\ 
{        } &20 &45.37   &3 &0.75 &1 &2.958 &5.77  &13.11 &$5.29\times10^{-7}$ & $2.89\times10^{-9}$  & $5.62\times10^{-10}$ & $2.94\times10^{-3}$ &$2.34\times10^{-9}$ \\ 
{        } &25 &56.72   &3 &0.75 &1 &3.059 &7.22  &16.39 &$5.29\times10^{-7}$ & $1.98\times10^{-9}$  & $3.85\times10^{-10}$ & $4.36\times10^{-3}$ &$1.53\times10^{-9}$ \\  
{        } &30 &68.06   &3 &0.75 &1 &3.132 &8.66  &19.66 &$5.29\times10^{-7}$ & $1.44\times10^{-9}$  & $2.80\times10^{-10}$ & $6.08\times10^{-3}$ &$1.10\times10^{-9}$ \\ 
{        } &35 &79.40   &3 &0.75 &1 &3.185 &10.11 &22.94 &$5.29\times10^{-7}$ & $1.09\times10^{-9}$  & $2.13\times10^{-10}$ & $8.11\times10^{-3}$ &$8.42\times10^{-10}$ \\ 
{        } &40 &90.75   &3 &0.75 &1 &3.226 &11.55 &26.22 &$5.29\times10^{-7}$ & $8.61\times10^{-10}$ & $1.67\times10^{-10}$ & $1.05\times10^{-2}$ &$6.69\times10^{-10}$ \\ 
{        } &45 &102.09  &3 &0.75 &1 &3.259 &13.0  &29.5  &$5.29\times10^{-7}$ & $6.94\times10^{-10}$ & $1.35\times10^{-10}$ & $1.31\times10^{-2}$ &$5.48\times10^{-10}$ \\  
{Model II} &50 &113.43  &3 &0.75 &1 &3.286 &14.44 &32.78 &$5.29\times10^{-7}$ & $5.71\times10^{-10}$ & $1.11\times10^{-10}$ & $1.61\times10^{-2}$ &$4.59\times10^{-10}$ \\ 
{        } &55 &124.78  &3 &0.75 &1 &3.308 &15.89 &36.05 &$5.29\times10^{-7}$ & $4.78\times10^{-10}$ & $9.29\times10^{-11}$ & $1.94\times10^{-2}$ &$3.93\times10^{-10}$ \\ 
{        } &60 &136.12  &3 &0.75 &1 &3.327 &17.33 &39.33 &$5.29\times10^{-7}$ & $4.07\times10^{-10}$ & $7.90\times10^{-11}$ & $2.31\times10^{-2}$ &$3.40\times10^{-10}$ \\ 
{        } &65 &147.46  &3 &0.75 &1 &3.343 &18.78 &42.61 &$5.29\times10^{-7}$ & $3.49\times10^{-10}$ & $6.79\times10^{-11}$ & $2.70\times10^{-2}$ &$2.98\times10^{-10}$ \\ 
{        } &70 &158.81  &3 &0.75 &1 &3.357 &20.22 &45.89 &$5.29\times10^{-7}$ & $3.04\times10^{-10}$ & $5.91\times10^{-11}$ & $3.13\times10^{-2}$ &$2.65\times10^{-10}$ \\ 
{        } &75 &170.15  &3 &0.75 &1 &3.369 &21.67 &49.17 &$5.29\times10^{-7}$ & $2.67\times10^{-10}$ & $5.19\times10^{-11}$ & $3.59\times10^{-2}$ &$2.37\times10^{-10}$ \\ 
{        } &80 &181.49  &3 &0.75 &1 &3.379 &23.11 &52.44 &$5.29\times10^{-7}$ & $2.36\times10^{-10}$ & $4.59\times10^{-11}$ & $4.08\times10^{-2}$ &$2.14\times10^{-10}$ \\ 
{        } &85 &192.83  &3 &0.75 &1 &3.389 &24.56 &55.72 &$5.29\times10^{-7}$ & $2.10\times10^{-10}$ & $4.08\times10^{-11}$ & $4.60\times10^{-2}$ &$1.94\times10^{-10}$ \\ 
{        } &90 &204.18  &3 &0.75 &1 &3.398 &26.0 &59.0 &$5.29\times10^{-7}$   & $1.88\times10^{-10}$ & $3.66\times10^{-11}$ & $5.16\times10^{-2}$ &$1.77\times10^{-10}$ \\ 
{        } &95 &215.52  &3 &0.75 &1 &3.405 &27.45 &62.28 &$5.29\times10^{-7}$ & $1.69\times10^{-10}$ & $3.30\times10^{-11}$ & $5.75\times10^{-2}$ &$1.63\times10^{-10}$ \\ 
{        } &100&226.87  &3 &0.75 &1 &3.412 &28.89 &65.56 &$5.29\times10^{-7}$ & $1.52\times10^{-10}$ & $2.99\times10^{-11}$ & $6.37\times10^{-2}$ &$1.50\times10^{-10}$ \\
\enddata
\tablecomments{
Column 1: model; 
Column 2: initial orbital separation ($AU$); 
Column 3: final orbital separation ($AU$); 
Column 4: initial mass of the stellar donor ($M_{\odot}$); 
Column 5: final mass of the stellar donor ($M_{\odot}$); 
Column 6: secondary mass [$M_{\odot}$]; 
Column 7: Bondi-Hoyle accretion radius ($AU$); 
Column 8: initial outer boundary of the disk ($AU$); 
Column 9: final outer boundary of the disk ($AU$); 
Column 10: mass-loss rate from the primary star ($M_{\odot} yr^{-1}$); 
Column 11: initial accretion rate into the disk ($M_{\odot} yr^{-1}$); 
Column 12: final accretion rate into the disk ($M_{\odot} yr^{-1}$); 
Column 13: final mass of the disk ($M_{\odot}$); 
Column 14: final accretion rate onto the star ($M_{\odot} yr^{-1}$);
}
\end{deluxetable*}
\begin{deluxetable*}{lcc|cc|c|c|cc|c|cc|c|c}
\label{table:results2}
\tabletypesize{\scriptsize}
\tablewidth{0pt} 
\tablecaption{Table of results model III and model IV. \label{tab:results}}
\colnumbers
\tablehead{
\colhead{Model} & 
\colhead{a$^{in}$} & 
\colhead{a$^{fin}$} & 
\colhead{ M$_1^{in}$} & 
\colhead{ M$_1^{fin}$} & 
\colhead{ M$_2$} & 
\colhead{ R$_a$} &
\colhead{ R$_{out}^{in}$} &
\colhead{ R$_{out}^{fin}$} &
\colhead{ $\dot{M}_{loss}$} & 
\colhead{ $\dot{M}_{disk}^{in}$} & 
\colhead{ $\dot{M}_{disk}^{fin}$} & 
\colhead{ $M_{disk}$} & 
\colhead{$\dot{M}_{in}$}
} 
\startdata 
{        } &10 &29.71  &5 &0.99 &1 &2.040 &2.51  &7.47  &$2.81\times10^{-6}$ &$2.93\times10^{-8}$ & $3.31\times10^{-9}$  & $2.12\times10^{-3}$ & $1.73\times10^{-8}$  \\ 
{        } &15 &44.57  &5 &0.99 &1 &2.377 &3.77  &11.21 &$2.81\times10^{-6}$ &$1.76\times10^{-8}$ & $1.99\times10^{-9}$  & $3.75\times10^{-3}$ & $6.49\times10^{-9}$  \\ 
{        } &20 &59.42  &5 &0.99 &1 &2.591 &5.03  &14.95 &$2.81\times10^{-6}$ &$1.18\times10^{-8}$ & $1.34\times10^{-9}$  & $5.63\times10^{-3}$ & $3.41\times10^{-9}$  \\ 
{        } &25 &74.27  &5 &0.99 &1 &2.738 &6.29  &18.69 &$2.81\times10^{-6}$ &$8.43\times10^{-9}$ & $9.55\times10^{-10}$ & $7.84\times10^{-3}$ & $2.14\times10^{-9}$  \\  
{        } &30 &89.13  &5 &0.99 &1 &2.847 &7.55  &22.43 &$2.81\times10^{-6}$ &$6.33\times10^{-9}$ & $7.17\times10^{-10}$ & $1.04\times10^{-2}$ & $1.49\times10^{-9}$  \\ 
{        } &35 &103.99 &5 &0.99 &1 &2.929 &8.81  &26.17 &$2.81\times10^{-6}$ &$4.92\times10^{-9}$ & $5.58\times10^{-10}$ & $1.33\times10^{-2}$ & $1.11\times10^{-9}$  \\ 
{        } &40 &118.83 &5 &0.99 &1 &2.995 &10.07 &29.91 &$2.81\times10^{-6}$ &$3.94\times10^{-9}$ & $4.46\times10^{-10}$ & $1.67\times10^{-2}$ & $8.68\times10^{-10}$ \\ 
{        } &45 &133.69 &5 &0.99 &1 &3.048 &11.32 &33.65 &$2.81\times10^{-6}$ &$3.22\times10^{-9}$ & $3.65\times10^{-10}$ & $2.04\times10^{-2}$ & $7.04\times10^{-10}$ \\  
{Model III}&50 &148.54 &5 &0.99 &1 &3.092 &12.58 &37.39 &$2.81\times10^{-6}$ &$2.69\times10^{-9}$ & $3.04\times10^{-10}$ & $2.46\times10^{-2}$ & $5.85\times10^{-10}$ \\ 
{        } &55 &163.39 &5 &0.99 &1 &3.128 &13.84 &41.13 &$2.81\times10^{-6}$ &$2.27\times10^{-9}$ & $2.58\times10^{-10}$ & $2.91\times10^{-2}$ & $4.97\times10^{-10}$ \\ 
{        } &60 &178.26 &5 &0.99 &1 &3.159 &15.1  &44.87 &$2.81\times10^{-6}$ &$1.95\times10^{-9}$ & $2.21\times10^{-10}$ & $3.41\times10^{-2}$ & $4.29\times10^{-10}$ \\ 
{        } &65 &193.09 &5 &0.99 &1 &3.186 &16.36 &48.61 &$2.81\times10^{-6}$ &$1.69\times10^{-9}$ & $1.91\times10^{-10}$ & $3.94\times10^{-2}$ & $3.75\times10^{-10}$ \\ 
{        } &70 &207.95 &5 &0.99 &1 &3.209 &17.62 &52.34 &$2.81\times10^{-6}$ &$1.47\times10^{-9}$ & $1.67\times10^{-10}$ & $4.51\times10^{-2}$ & $3.32\times10^{-10}$ \\ 
{        } &75 &222.81 &5 &0.99 &1 &3.230 &18.88 &56.09 &$2.81\times10^{-6}$ &$1.30\times10^{-9}$ & $1.48\times10^{-10}$ & $5.10\times10^{-2}$ & $2.96\times10^{-10}$ \\ 
{        } &80 &237.68 &5 &0.99 &1 &3.248 &20.14 &59.83 &$2.81\times10^{-6}$ &$1.16\times10^{-9}$ & $1.31\times10^{-10}$ & $5.71\times10^{-2}$ & $2.66\times10^{-10}$ \\ 
{        } &85 &252.48 &5 &0.99 &1 &3.265 &21.4  &63.56 &$2.81\times10^{-6}$ &$1.04\times10^{-9}$ & $1.17\times10^{-10}$ & $6.34\times10^{-2}$ & $2.41\times10^{-10}$ \\ 
{        } &90 &267.34 &5 &0.99 &1 &3.279 &22.65 &67.3  &$2.81\times10^{-6}$ &$9.33\times10^{-10}$ & $1.06\times10^{-10}$ & $6.97\times10^{-2}$& $2.20\times10^{-10}$ \\ 
{        } &95 &282.24 &5 &0.99 &1 &3.292 &23.91 &71.05 &$2.81\times10^{-6}$ &$8.44\times10^{-10}$ & $9.56\times10^{-11}$ & $7.62\times10^{-2}$& $2.02\times10^{-10}$ \\ 
{        } &100&297.06 &5 &0.99 &1 &3.304 &25.17 &74.78 &$2.81\times10^{-6}$ &$7.68\times10^{-10}$ & $8.69\times10^{-11}$ & $8.28\times10^{-2}$& $1.86\times10^{-10}$ \\
\hline
{        } &10 &35.75  &8 &1.44 &1 &1.57 &2.20 & 7.88  & $8.97\times10^{-6}$ & $5.52\times10^{-8}$ & $4.32\times10^{-9}$  & $3.68\times10^{-3}$ &$2.87\times10^{-8}$ \\ 
{        } &15 &53.63  &8 &1.44 &1 &1.93 &3.30 & 11.82 & $8.97\times10^{-6}$ & $3.70\times10^{-8}$ & $2.89\times10^{-9}$  & $6.37\times10^{-3}$ &$1.02\times10^{-8}$ \\ 
{        } &20 &71.51  &8 &1.44 &1 &2.18 &4.41 & 15.76 & $8.97\times10^{-6}$ & $2.65\times10^{-8}$ & $2.08\times10^{-9}$  & $9.39\times10^{-3}$ &$5.12\times10^{-9}$ \\ 
{        } &25 &89.38  &8 &1.44 &1 &2.36 &5.51 & 19.71 & $8.97\times10^{-6}$ & $1.99\times10^{-8}$ & $1.56\times10^{-9}$  & $1.28\times10^{-2}$ &$3.08\times10^{-9}$ \\  
{        } &30 &107.26  &8 &1.44 &1 &2.49 &6.61& 23.65 & $8.97\times10^{-6}$ & $1.55\times10^{-8}$ & $1.22\times10^{-9}$  & $1.67\times10^{-2}$ &$2.08\times10^{-9}$ \\ 
{        } &35 &125.13  &8 &1.44 &1 &2.61 &7.71 &27.59 & $8.97\times10^{-6}$ & $1.24\times10^{-8}$ & $9.74\times10^{-10}$ & $2.11\times10^{-2}$ &$1.52\times10^{-9}$ \\ 
{        } &40 &142.99  &8 &1.44 &1 &2.69 &8.82 &31.53 & $8.97\times10^{-6}$ & $1.02\times10^{-8}$ & $7.98\times10^{-10}$ & $2.56\times10^{-2}$ &$1.16\times10^{-9}$ \\ 
{        } &45 &160.89  &8 &1.44 &1 &2.77 &9.92 &35.48 & $8.97\times10^{-6}$ & $8.50\times10^{-9}$ & $6.65\times10^{-10}$ & $3.03\times10^{-2}$ &$9.26\times10^{-10}$ \\ 
{Model IV} &50 &178.76  &8 &1.44 &1 &2.83 &11.02&39.42 & $8.97\times10^{-6}$ & $7.20\times10^{-9}$ & $5.63\times10^{-10}$ & $3.48\times10^{-2}$ &$7.61\times10^{-10}$ \\ 
{        } &55 &196.61  &8 &1.44 &1 &2.88 &12.13 &43.35& $8.97\times10^{-6}$ & $6.18\times10^{-9}$ & $4.83\times10^{-10}$ & $3.93\times10^{-2}$ &$6.39\times10^{-10}$ \\ 
{        } &60 &214.47  &8 &1.44 &1 &2.93 &13.23 &47.29& $8.97\times10^{-6}$ & $5.35\times10^{-9}$ & $4.19\times10^{-10}$ & $4.37\times10^{-2}$ &$5.48\times10^{-10}$ \\ 
{        } &65 &232.36  &8 &1.44 &1 &2.97 &14.33 &51.24& $8.97\times10^{-6}$ & $4.69\times10^{-9}$ & $3.68\times10^{-10}$ & $4.79\times10^{-2}$ &$4.76\times10^{-10}$ \\  
{        } &70 &250.25  &8 &1.44 &1 &3.01 &15.43 &55.18& $8.97\times10^{-6}$ & $4.14\times10^{-9}$ & $3.23\times10^{-10}$ & $5.22\times10^{-2}$ &$4.19\times10^{-10}$ \\ 
{        } &75 &268.12  &8 &1.44 &1 &3.04 &16.54 &59.13& $8.97\times10^{-6}$ & $3.68\times10^{-9}$ & $2.88\times10^{-10}$ & $5.64\times10^{-2}$ &$3.72\times10^{-10}$ \\ 
{        } &80 &285.95  &8 &1.44 &1 &3.06 &17.64 &63.06& $8.97\times10^{-6}$ & $3.29\times10^{-9}$ & $2.58\times10^{-10}$ & $6.05\times10^{-2}$ &$3.34\times10^{-10}$ \\ 
{        } &85 &303.85  &8 &1.44 &1 &3.09 &18.74 &67.0 & $8.97\times10^{-6}$ & $2.96\times10^{-9}$ & $2.32\times10^{-10}$ & $6.46\times10^{-2}$ &$3.02\times10^{-10}$ \\ 
{        } &90 &321.77  &8 &1.44 &1 &3.11 &19.85 &70.96&$8.97\times10^{-6}$  & $2.68\times10^{-9}$ & $2.09\times10^{-10}$ & $6.88\times10^{-2}$ &$2.75\times10^{-10}$ \\ 
{        } &95 &339.49  &8 &1.44 &1 &3.13 &20.95 &74.87 &$8.97\times10^{-6}$ & $2.44\times10^{-9}$ & $1.91\times10^{-10}$ & $7.29\times10^{-2}$ &$2.52\times10^{-10}$ \\ 
{        } &100 &357.33 &8 &1.44 &1 &3.15 &22.05 &78.8  &$8.97\times10^{-6}$ & $2.23\times10^{-9}$ & $1.74\times10^{-10}$ & $7.43\times10^{-2}$ &$2.31\times10^{-10}$ \\
\enddata
\tablecomments{
Column 1: model; 
Column 2: initial orbital separation ($AU$); 
Column 3: final orbital separation ($AU$); 
Column 4: initial mass of the stellar donor ($M_{\odot}$); 
Column 5: final mass of the stellar donor ($M_{\odot}$); 
Column 6: secondary mass [$M_{\odot}$]; 
Column 7: Bondi-Hoyle accretion radius ($AU$); 
Column 8: initial outer boundary of the disk ($AU$); 
Column 9: final outer boundary of the disk ($AU$); 
Column 10: mass-loss rate from the primary star ($M_{\odot} yr^{-1}$); 
Column 11: initial accretion rate into the disk ($M_{\odot} yr^{-1}$); 
Column 12: final accretion rate into the disk ($M_{\odot} yr^{-1}$); 
Column 13: final mass of the disk ($M_{\odot}$); 
Column 14: final accretion rate onto the star ($M_{\odot} yr^{-1}$);
}
\end{deluxetable*}
\begin{deluxetable*}{lcc|cc|c|c|cc|c|cc|c|l}
\label{table:results_mira}
\tabletypesize{\scriptsize}
\tablewidth{0pt} 
\tablecaption{Table of results for \textit{Mira} system model. \label{tab:results}}
\colnumbers
\tablehead{
\colhead{Model} & 
\colhead{a$^{in}$} & 
\colhead{a$^{fin}$} & 
\colhead{ M$_1^{in}$} & 
\colhead{ M$_1^{fin}$} & 
\colhead{ M$_2$} & 
\colhead{ R$_a$} &
\colhead{ R$_{out}^{in}$} &
\colhead{ R$_{out}^{fin}$} &
\colhead{ $\dot{M}_{loss}$} & 
\colhead{ $\dot{M}_{disk}^{in}$} & 
\colhead{ $\dot{M}_{disk}^{fin}$} & 
\colhead{ $M_{disk}$} & 
\colhead{$\dot{M}_{in}$}
} 
\startdata 
{Mira} &60.0 &76.13  &1.62 &1.18 &0.6 &2.062 &16.84 &21.47  &$4.40\times10^{-7}$ & $1.55\times10^{-10}$ & $9.51\times10^{-11}$ & $0.82\times10^{-3}$ & $1.03\times10^{-10}$\\ 
\enddata
\tablecomments{
Column 1: model; 
Column 2: initial orbital separation ($AU$); 
Column 3: final orbital separation ($AU$); 
Column 4: initial mass of the stellar donor ($M_{\odot}$); 
Column 5: final mass of the stellar donor ($M_{\odot}$); 
Column 6: secondary mass [$M_{\odot}$]; 
Column 7: Bondi-Hoyle accretion radius ($AU$); 
Column 8: initial outer boundary of the disk ($AU$); 
Column 9: final outer boundary of the disk ($AU$); 
Column 10: mass-loss rate from the primary star ($M_{\odot} yr^{-1}$); 
Column 11: initial accretion rate into the disk ($M_{\odot} yr^{-1}$); 
Column 12: final accretion rate into the disk ($M_{\odot} yr^{-1}$); 
Column 13: final mass of the disk ($M_{\odot}$); 
Column 14: final accretion rate onto the star ($M_{\odot} yr^{-1}$);
}
\end{deluxetable*}
%
\FloatBarrier
\nocite{*}

\bibliographystyle{aasjournal}

\end{document}